\documentclass[submission,copyright,creativecommons]{eptcs}
\providecommand{\event}{Foclasa 2015} 
\usepackage{breakurl}             

\usepackage{times}
\usepackage{caption}
\usepackage{paralist}
\usepackage{graphicx}
\usepackage{subfigure}
\usepackage{listings}
\usepackage{amsmath}
\usepackage{listings}
\usepackage{wrapfig}
\usepackage{moreverb}
\usepackage{url}
\usepackage[boxed]{algorithm}
\usepackage{algorithmic}
\usepackage[all,2cell,ps]{xy}
\usepackage{times}
\usepackage{paralist}
\usepackage{graphicx}
\usepackage{wasysym}
\usepackage{amssymb}
\usepackage{amsfonts}
\usepackage{moreverb}
\usepackage{url}
\usepackage{cite}
\usepackage{color}
\usepackage{lscape}
\usepackage{amsthm}
\newtheorem{example}{Example}
\newtheorem{definition}{Definition}
\newtheorem{lemma}{Lemma}
\newtheorem*{proof*}{Proof}
\newtheorem{theorem}{Theorem} 
\newcommand{\Nat}{{\mathbb N}}

\title{A Constraint-based Approach for Generating Transformation Patterns}
\author{Asma Cherif
\institute{Umm Al-Qura University\\ Makkah, Saudi Arabia} 
\email{ahcherif@mail.uqu.edu.sa}
\and
Abdessamad Imine
\institute{Lorraine Univeristy and Inria Nancy Grand-Est\\
Nancy, France}
\email{\quad imine@loria.fr }
}
\def\titlerunning{A Constraint-based Approach for Generating Transformation Patterns}
\def\authorrunning{A. Cherif, A. Imine }
\begin{document}\sloppy
\maketitle

\begin{abstract}

Undoing operations is an    indispensable feature for many
collaborative applications, mainly  collaborative editors. It
provides the ability to restore a correct state of the shared data
after erroneous operations. In particular, selective undo allows
users to undo any operation and is based on rearranging operations
in the history using the Operational Transformation (OT) approach. OT is an optimistic replication technique that allows many users to concurrently update the shared data and exchange
their updates in any order. To ensure consistency, OT enforces the out-of-order execution of concurrent updates using transformation functions that must have been planned in advance.
It is a challenging task how to meaningfully combine OT and undo approaches while preserving consistency. Indeed, undoing operations that are received and executed out-of-order at different collaborating sites inevitably leads to divergence cases.
Even though various   undo solutions have been proposed over
the recent years, they are either limited or erroneous. 

In this paper, we propose a constraint-based approach to address the undo problem that
is formulated as a Constraint Satisfaction Problem (CSP). 
By using CSP approach, we are able to analyze all covered transformation cases for coordinating collaborative objects with finite size of operations. This allows to devise undoable transformation patterns which considerably simplifies the design of collaborative objects. We also study the relation between commutativity and undoability which enables us to state
a very important theoretical result. Indeed, we prove that   commutativity is necessary and sufficient to achieve undoability for small sets of operations (of sizes 2 and 4) and only sufficient otherwise. This work represents a step forward toward
a practical use of CSP techniques for designing safe OT-based collaborative applications.
\end{abstract}
 

\section{Introduction}
\noindent\textbf{Motivation.}
Nowadays, collaborative applications are becoming more widespread
due to the powerful evolution of networks and their services. For instance, collaborative editors (e.g. Google Docs) allow several and dispersed users to simultaneously cooperate  with each other in order to manipulate a shared object (e.g. a multimedia document). To ensure availability of data as well as high local responsiveness, these applications resort to replicating shared objects.
So, the updates are applied in different orders
at different replicas of the object. This potentially leads to divergent (or different)
replicas, an undesirable situation for collaborative applications.
\emph{Operational Transformation} (OT) is an optimistic technique which has been proposed to overcome the
divergence problem~\cite{Ellis89,Sun98}. It
enforces to some extent the commutativity between conflicting
operations without using roll-back, but by using transformations that must have been planned in advance.
Indeed, the OT approach consists of application-dependent transformation algorithm $IT(op_1, op_2)$ to compute
the transformation of operation $op_1$  which is a  new variant of $op_1$ that will be executed after operation $op_2$. Thus, for every possible pair of concurrent
operations, the application programmer has to define in advance how to merge these operations regardless of execution
order. 
To ensure the convergence of all replicas, a transformation algorithm requires to   satisfy   two \emph{transformation
properties}~\cite{Ressel.ea:96}, namely $\mathbf{TP1}$ and $\mathbf{TP2}$ (see Section~\ref{sec:do}).
OT is used in many
collaborative editors including Joint Emacs~\cite{Ressel.ea:96}, CoWord~\cite{Sun06}, CoPowerPoint~\cite{Sun06}, the Google Wave \footnote{\url{http://www.waveprotocol.org/whitepapers/operational-transform}}, and Google Docs \footnote{\url{http://en.wikipedia.org/wiki/Google_Docs}}.

Undoing operations is an  indispensable feature for many collaborative applications, mainly real
time collaborative editors. It provides the ability to restore a correct state of the shared data after erroneous operations. In particular, \emph{selective undo} consists in undoing any operation from the local history of operations performed locally or received from  remote sites. It is especially required for  maintaining convergence in access control-based collaborative editors~\cite{CherifIR11,CherifIR14}. Indeed, in collaborative applications, operations are received out-of-order at different collaborating sites. Thus undoing an illegal operation at one site may necessitate to undo it in a different form (\textit{i.e.} its transformation form) at another according to its reception order. To correctly undo operations, three inverse properties,  namely $\mathbf{IP1}$, $\mathbf{IP2}$ and $\mathbf{IP3}$~\cite{Atul94,Sun02,SunS09,Ferrie04}  were proposed (see Section~\ref{sec:undo}). Combining OT and undo approaches while preserving data convergence
remains an open and challenging issue since many divergence cases may be encountered when undoing operations.
Even though many solutions were proposed over the recent years, designing undo schemes for collaborative applications is a hard task
since each proposed solution has either a limitation (i.e. it relaxes some constraints at the expense of system performance) or a counterexample showing it is not correct~\cite{SunS09, BinShao10}.

\medskip
\noindent\textbf{Contribution.}
In this paper, we present a theoretical study of the undoability problem in collaborative applications. For any shared object with a set of primitive operations, 
we provide a formal model to investigate the existence  of convergent transformation functions satisfying inverse properties. As approaching these transformation functions turns out to be combinatorial in nature, we resort to constraint programming  to formalize the  undoability problem as a Constraint Satisfaction Problem (CSP).
Thus, we define a collaborative application as a shared object whose state must satisfy both transformation and inverse properties. We use our model to devise transformation patterns that guarantee both the  convergence of shared data and the correctness of the undo approach. Furthermore, we study the relation between undoability and commutativity. Yet the OT approach was proposed to go beyond the commutativity, we prove that commutativity is necessary and sufficient to correctly undo operations in  consistent objects of size $2$ and $4$ and only sufficient otherwise.

\medskip
\noindent\textbf{Outline.} The paper is organized as follows: in Section 2, we present  OT approach by describing how to do and undo user updates.
Section 3 describes our formal model and shows how we formulate the undoability as a CSP. In Section 4, we study the undoable transformation functions provided by our solver. Section 5 discusses our results.  
We review related work in Section 6 and conclude the paper with future research in Section 7.

\section{Operational Transformation Approach}  
To get started, we first present the ingredients of OT approach by describing how to do and undo
user updates in collaborative context.  

\subsection{Doability of Updates}\label{sec:do}
OT is an optimistic replication  technique which allows many users (or
sites) to generate operations in order to concurrently modify the  shared
data and next to coordinate
their divergent replicas in order to obtain the same  data~\cite{Ellis89,Sun98}. The operations
of each  site are  executed on the  local replica  immediately without
being blocked or delayed, and then are propagated to other sites to be
executed again.  Accordingly, every operation is processed in four steps:
\begin{inparaenum}[(i)]
\item \textit{generation} on one site;
\item \textit{broadcast} to other sites;
\item \textit{reception} on one site;
\item \textit{execution} on one site.
\end{inparaenum}

As any distributed application, exchanging operations requires to track relations
between these operations. Two relations are often given in the
literature~\cite{Ellis89,Sun98}:
\vspace{-0.1cm}
\begin{definition}\textbf{\emph{(Causality  and Concurrency Relations).}}\label{Def:caus}
Let an operation $op_1$ be generated at site $i$ and an  operation $op_2$
be generated at site $j$. We say that $op_2$ \emph{causally depends} on
$op_1$, denoted $op_1 \rightarrow op_2$, iff:
\begin{inparaenum}[(i)]
\item $i=j$ and $op_1$ was generated before $op_2$; or,
\item $i\neq j$ and the execution of $op_1$ at site $j$ has happened before
      the generation of $op_2$.
\end{inparaenum}
Two operations $op_1$ and $op_2$ are said to be \emph{concurrent},
denoted by $op_1 \parallel op_2$, iff neither $op_1 \rightarrow op_2$ nor
$op_2 \rightarrow op_1$.
\end{definition}

As a long established convention in OT-based
collaborative applications~\cite{Ellis89, Sun.ea:98}
, the
\emph{timestamp vectors} are used to determine the causality and
concurrency relations between operations.
Due to high  communication latencies in wide-area  and  mobile
wireless  networks the replication of collaborative objects is commonly
used in distributed collaborative systems. But this choice is not
without problem. Indeed, one  of the  significant  issues when  building collaborative
editors with a replicated  architecture and an arbitrary communication
of messages between users  is the \textit{consistency maintenance} (or
\textit{convergence})  of all  replicas. To  illustrate  this problem,
we give the following example:

\begin{example}\label{exmp:upDown}
Consider a shared binary register where two primitive operations modify the state of a bite from $0$ to $1$ and vice versa:
\begin{inparaenum}[(i)]
\item $Up$ to turn on the register;
\item $Down$ to turn off the register.
\end{inparaenum}
Suppose that this shared register is manipulated concurrently by two users, as depicted in
Figure~\ref{fig:div}(a). Initially, both copies of the shared register contain $0$.
User   $1$  executes  operation  $op_1 =  Up$  to turn the local state to $1$.  Concurrently, user  $2$   performs  $op_2  =  Down$.   When  $op_1$  is  received  and
executed   on   site   $2$,   it  produces   the   expected   state $1$.  But,  at  site $1$, $op_2$
does not take  into account that $op_1$ has  been executed before it
  and it produces the  state $0$.  Thus, the final state at site
  $1$  is different  from that of  site $2$.

\begin{figure}[h!]
\begin{minipage}[t]{0.45\linewidth}
\centerline{\includegraphics[scale=0.3]{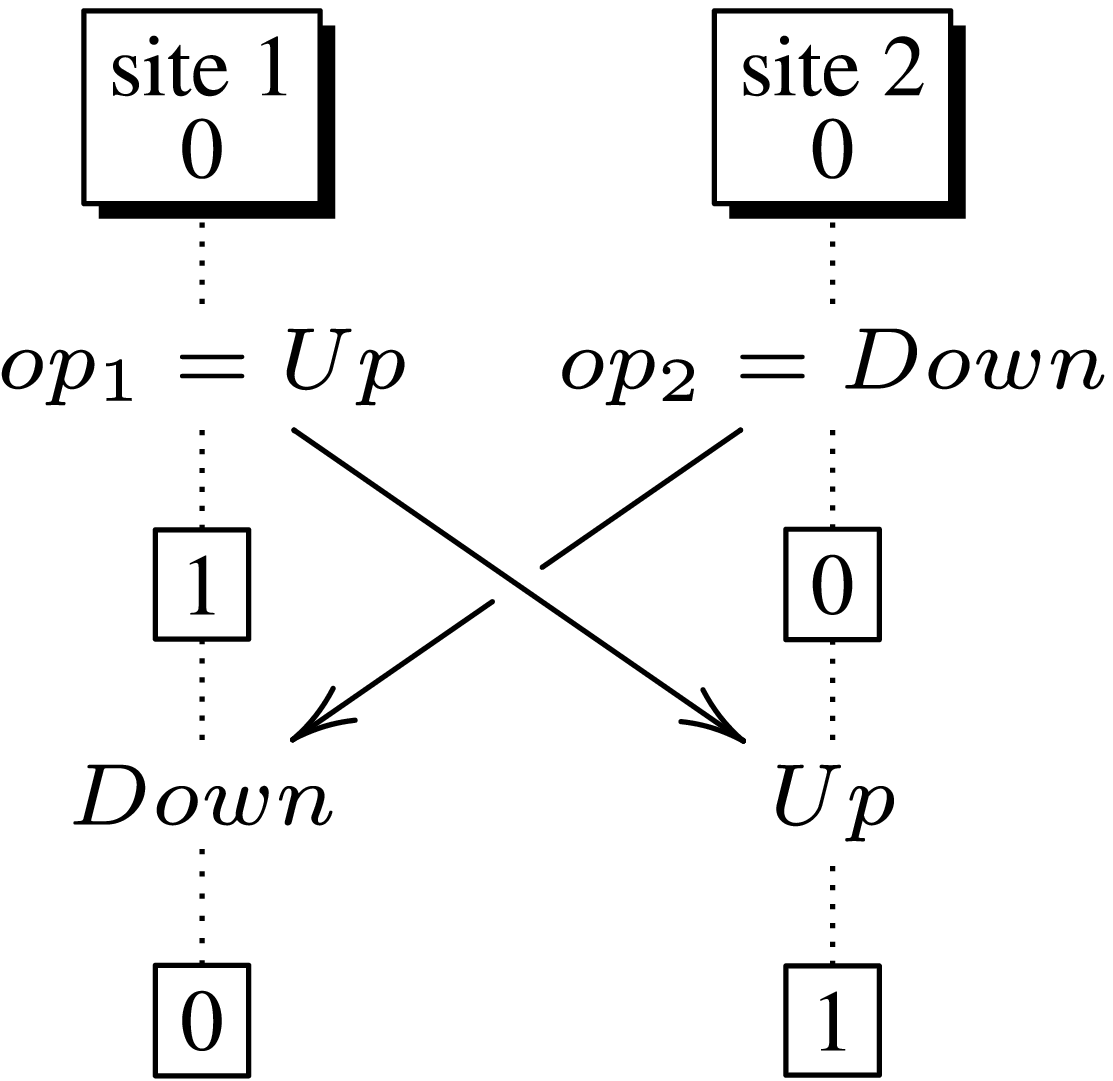}}
\centerline{(a) Incorrect integration.}
 \end{minipage}
 \hspace{.75cm}
 \begin{minipage}[t]{0.45\linewidth}
\centerline{\includegraphics[scale=0.3]{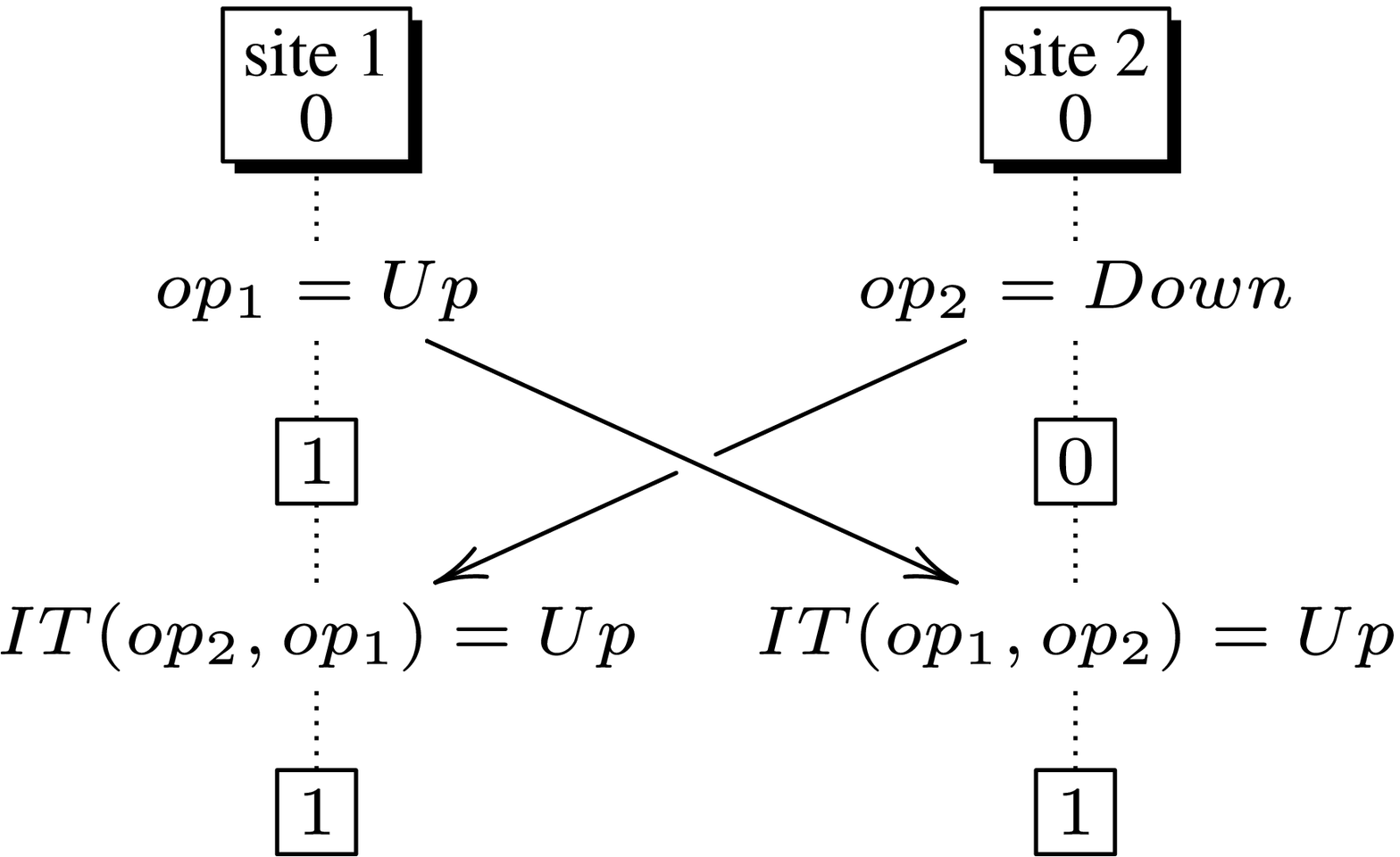}}
\centerline{(b) Correct integration.}
 \end{minipage}
\caption{Serialization of concurrent updates}\label{fig:div}
\vspace{-5mm}
\end{figure}

\end{example}

To maintain convergence, a serialization by transformation can be used. When  user $1$  gets an operation $op$ that
was previously executed  by  user $2$ on  his replica of the
shared  object,   he  does  not  necessarily  integrate $op$  by
executing it  ``as is''   on  his replica. He will rather  execute a
variant  of $op$,  denoted by  $op'$ (called  a \emph{transformation} of
$op$) that  \textit{intuitively intends to achieve the  same effect as
  $op$}.
 To do this, OT has been proposed to provide
application-dependent  transformation algorithm called
\emph{Inclusion Transformation} $IT$ such  that for  every
possible pair of concurrent operations, the application programmer has to
specify  how to  merge these  operations regardless  of  reception 
order~\cite{Ellis89,Sun98,Sun.ea:98}.
For instance, the following transformation function gives \textit{transformation cases} for the set operations $\{Up,Down\}$ given in Example~\ref{exmp:upDown}.
\begin{small}
\begin{lstlisting}
$IT_{bin}(op_1,op_2):op'_1$
   Choice of $op_1$ and $op_2$
      Case ($op_1=Up$ and $op_2=Up$):
                $op'_1 \leftarrow Up$
      Case ($op_1=Up$ and $op_2=Down$):
                $op'_1 \leftarrow Up$
      Case ($op_1=down$ and $op_2=Up$):
                $op'_1 \leftarrow Up$
      Case ($op_1=Down$ and $op_2=Down$):
                $op'_1 \leftarrow Down$
   End
\end{lstlisting}
\end{small}

\medskip
In the following, we show how to correctly merge operations using the previous algorithm. 
\begin{example}\label{exmp:e12}
In Figure~\ref{fig:div}(b),  we illustrate the effect of  $IT$ on the
previous example. At site $1$, $op_2$ needs
to    be    transformed   in order to include the effects of $op_1$:
$op'_2=IT(Down,Up) =Up$.  The  operation $Down$ is transformed to $Up$ since another  $Up$ was concurrently generated.
\end{example}

\noindent\textbf{OT Properties.} 
OT approach requires that every site stores all executed operations in a buffer also called a \emph{log}. A log is a sequence of operations buffered at their execution order.
Notation $op_1\cdot op_2\cdot \ldots\cdot op_n$ represents the operation sequence of $n$ operations.
By abuse of notation, we denote by $st\cdot X = st'$ the operation (or an operation sequence)
$X$ that is executed
on a replica state $st$ and produces a replica state $st'$.  We can define an equivalence
relation between operation sequences as follows:
\begin{definition}\textbf{\emph{(Equivalence between sequences of operations).}}\label{Def:tequiv}
Two sequences $seq_1$ and $seq_2$ are \emph{equivalent }, denoted by
$seq_1 \equiv seq_2$, iff  
 $seq_1$ and $seq_2$ produce the same state, \textit{i.e.} $st\cdot seq_1 = st\cdot seq_2$ for every state $st$. 
\end{definition}

Transforming  any operation  $op$ against a sequence of operations  $seq$ is  denoted by
$IT^*(op,seq)$ and is recursively defined as follows:
\begin{itemize}\vspace{-1mm}
\item $IT^*(op,\emptyset)=op\mbox{ where } \emptyset \mbox{ is the empty sequence;}$\vspace{-2mm}
\item $IT^*(op,op_1\cdot op_2\cdot\ldots\cdot op_n)\, =\, IT^*(IT(op,op_1),op_2\cdot\ldots\cdot op_n)$
\end{itemize}\vspace{-1mm}
We say that $op$ has been concurrently generated according to all
operations of $seq$.

To ensure the convergence of all replicas, a transformation algorithm requires to   satisfy   two
properties~\cite{Ressel.ea:96,Sun02}, called \emph{transformation properties}.

\begin{definition}\textbf{\emph{(Transformation properties).}}\label{Def:TP1TP2}
Let $IT$ be an inclusion transformation function. For all   $op_1$, $op_2$ and $op_3$
pairwise concurrent operations, $IT$ is \emph{correct} iff the following properties are
satisfied: 
\begin{itemize}
\item \textbf{Property $\mathbf{TP1}$}: $st\cdot op_1\cdot IT(op_2,op_1)\,=\,
      st\cdot op_2\cdot IT(op_1,op_2)$, for every state $st$.
\item \textbf{Property $\mathbf{TP2}$}:
        $IT^*(op_3,op_1\cdot IT(op_2,op_1))\,=\,IT^*(op_3,op_2\cdot IT(op_1,op_2))$.

\end{itemize}
\end{definition}

$\mathbf{TP1}$ defines a  \emph{state identity} and ensures that  if $op_1$ and
$op_2$ are concurrent, the effect of executing $op_1$ before $op_2$ is
the  same  as  executing  $op_2$  before  $op_1$.  This  condition  is
necessary but not sufficient  when the number of concurrent operations
is greater than two.  As  for $\mathbf{TP2}$, it ensures that transforming $op_3$
against equivalent  operation sequences  results in  the same operation.

Properties $\mathbf{TP1}$  and $\mathbf{TP2}$ are  sufficient to ensure  the convergence
for \textit{any number} of concurrent operations which can be
executed in \textit{arbitrary order}~\cite{Ressel.ea:96,Lus03}. 
Moreover, based on transformation properties, we can reorder operations in a sequence without altering
the resulting state of the original sequence which is very useful for undoing concurrent operations.

In the following, we say that a transformation function $IT$ is correct if it verifies both properties $\mathbf{TP1}$ and $\mathbf{TP2}$. 
For instance, the    function $IT_{bin}$ presented earlier  is correct since it verifies both properties.

\subsection{Undoability of Updates}\label{sec:undo}
\begin{figure}[!h]
\vspace{-.5cm}
\begin{scriptsize}
\centerline{\includegraphics[scale=0.3]{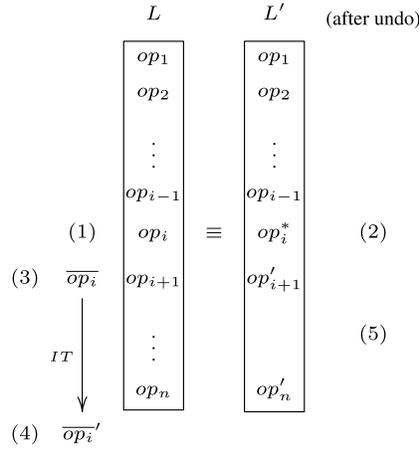}}
\end{scriptsize}
\vspace{-.2cm}
\caption{Undo Scheme.}\label{fig:undoScheme}
\end{figure}

The ability to undo operations performed by a user   is a   very useful feature allowing to reverse erroneous operations. Thus, it is possible to restore a previous convergent state without being obliged to redo all the work performed on a document.
  The selective \emph{undo} mechanism allows for maintaining convergence in access control-based collaborative editors~\cite{CherifIR11}. Indeed, in such applications any operation may be dynamically  revoked even if it is already executed. So, enforcing a dynamic access control policy requires to selectively undo operations from a given log. This approach  is based on rearranging operations in the history using the  OT approach. 
Consequently, it is primordial to log all executed operations to accomplish an undo scheme. Furthermore, all operations should be undoable. For this, we suppose that each operation $op$ has an inverse denoted by $\overline{op}$. As proposed in~\cite{Atul94,Sun02}, to
selectively undo operation $op_i$ from a given log say $L=op_1 \cdot op_2 \cdot \ldots \cdot op_i \cdot \ldots \cdot op_n$, we proceed by the following consecutive steps as illustrated in Figure~\ref{fig:undoScheme}:
\begin{enumerate}[(1)]
  \item Find $op_i$ in $L$;\vspace{-1.5mm}
  \item Mark $op_i$ as an undone operation: $op^*_i$;\vspace{-1.5mm}
  \item Generate $\overline{op_i}$;\vspace{-1.5mm}
  \item Calculate $\overline{op}'=IT^*(\overline{op_i}, op_{i+1} \cdot \ldots \cdot op_n)$ that integrates the effect of the sequence following $op_i$ in $L$;\vspace{-1.5mm}
  \item Exclude\footnote{We can exclude the effect of $op_i$ from the sublog $op_{i+1} \cdot \ldots \cdot op_n$ using this small algorithm:\\
  $op\leftarrow \overline{op_i}$\\
 \textbf{for} $j$ from $i+1$ to $n$ \textbf{do}\\
 $op'_{j} \leftarrow IT(op_j,op)$\\
 $op \leftarrow IT(op,op_j)$\\
 \textbf{end for}
  } the effect of $op_i$ from the log by including the effect of $\overline{op_i}$ inside the sequence
  $op_{i+1} \cdot \ldots \cdot op_n$ (\textit{i.e.} the sequence following $op^*_i$). 
  The resulting sequence   is then $op'_{i+1} \cdot \ldots \cdot op'_n$;\vspace{-1.5mm}
  \item Execute $\overline{op}'$.
\end{enumerate}
 
Finally, the sequence $L\cdot \overline{op_i}'$ should be equivalent to $L'$ so that the undoability is correct.


\medskip\noindent\textbf{Undo Properties. }Three inverse properties $\mathbf{IP1}$, $\mathbf{IP2}$ and $\mathbf{IP3}$, have been proposed in the literature~\cite{Atul94,Sun02,SunS09,Ferrie04} to formalize the correctness of a transformation-based undo scheme.

\begin{definition}[Inverse Property 1 ($\mathbf{IP1}$)]
\label{Def:IP1}
Given  any operation $op$ and its inverse $\overline{op}$, then:
$ op\cdot \overline{op}\equiv \emptyset$.
\end{definition}

Property $\mathbf{IP1}$ means the operation sequence $op \cdot \overline{op}$ must not affect  the object state and is not related to transformation functions.

\begin{definition}[Inverse Property 2 ($\mathbf{IP2}$)]
\label{Def:IP2}
Given a correct transformation function $IT$ and any two  operations $op_1$ and $op_2$ then:
$IT(IT(op_1,op_2), \overline{op_2})= op_1$.
\end{definition}

As the sequence $op_2 \cdot \overline{op_2}$  have no effect, property $\mathbf{IP2}$ means transforming $op_1$
against $op_2$ and its inverse $\overline{op_2}$ must result in the same operation.

\begin{definition}[Inverse Property 3 ($\mathbf{IP3}$)]
\label{Def:IP3}
Given a transformation function $IT$ and any two  concurrent operations $op_1$ and $op_2$ with
$op'_1=IT(op_1,op_2)$ and $op'_2=IT(op_2,op_1)$. If sequences $ op_1\cdot op'_2 \equiv op_2\cdot op'_1$
 then
$IT(\overline{op_1}, op'_2) = \overline{IT(op_1,op_2)}$.
\end{definition}

Property $\mathbf{IP3}$ means that the operation executed to undo $op_1$ in $op_1\cdot op'_2$ is the same as the operation executed
to undo it transformed form $op'_1$ in $op_2\cdot op'_1$.

The violation of one of the previous three properties, leads to divergence situations referred to as \emph{puzzles}. This is due to the fact that even though the considered transformation functions are correct (\textit{i.e.},
they satisfy the transformation properties $\mathbf{TP1}$ and $\mathbf{TP2}$) they are not sufficient to preserve the data
convergence when undoing operations.
Puzzles are subtle but characteristic scenarios allowing to conceive a correct undo solution.
All known undo puzzles are  due to the
violation of $IP2$ or $IP3$  by transformation functions.
For instance, in group editors, trying to identify and solve various puzzles has been a major stimulus in developing and verifying various novel collaborative editing techniques~\cite{Sun98, Ressel99, Sun00}. The ability to solve identified undo puzzles is a necessary condition and an important indicator of the
soundness of an undo solution.  

\begin{figure}[htbp]
\begin{minipage}[t]{0.50\linewidth}
\begin{tiny}
\centerline{\includegraphics[scale=0.3]{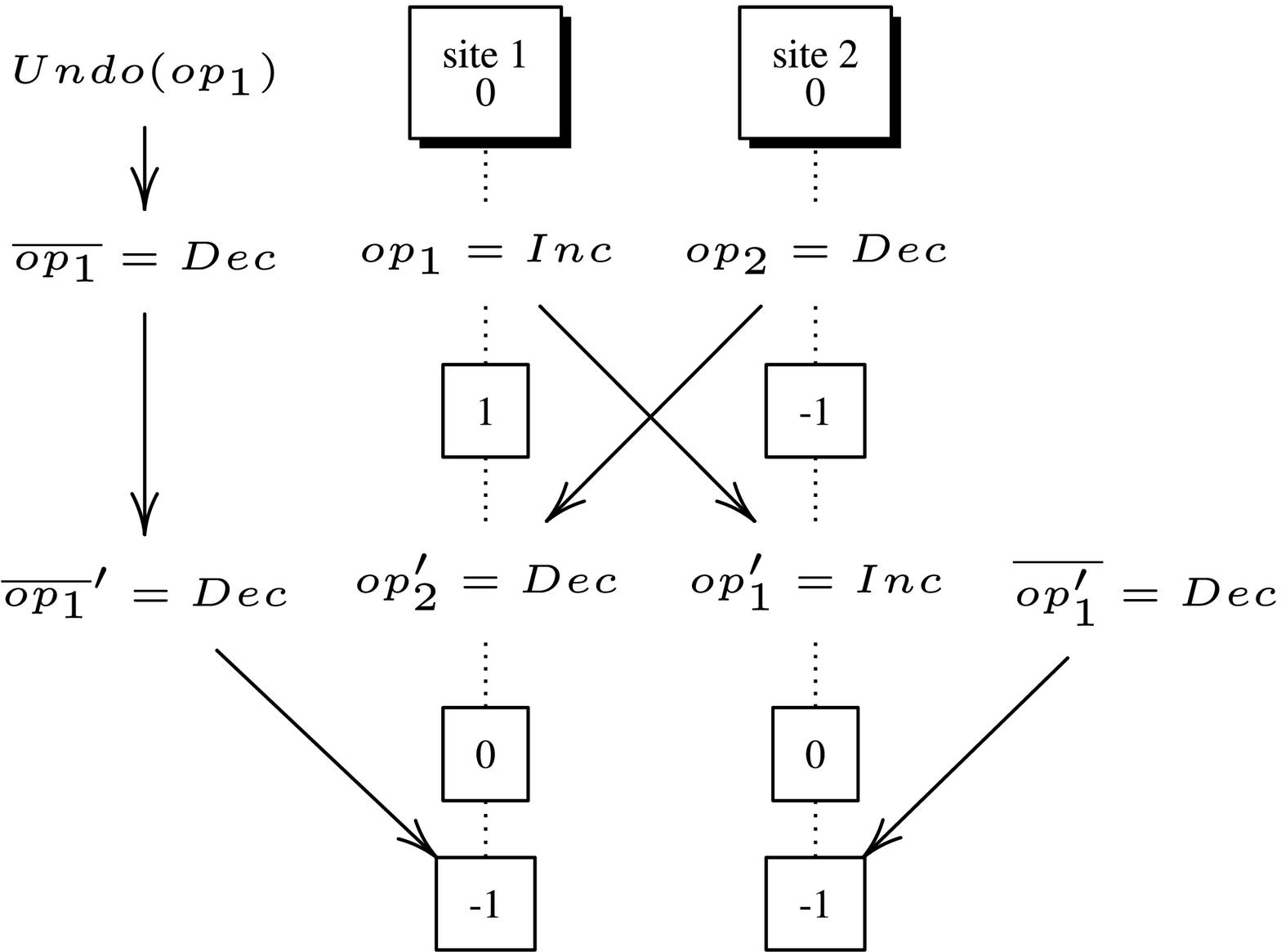}}
\end{tiny}
\begin{small}
\centerline{(a) $\mathbf{IP3}$ preservation for a scenario of $\{Inc(), Dec()\}$.} 
\end{small}
\end{minipage}
\begin{minipage}[t]{0.50\linewidth}
\centerline{\includegraphics[scale=0.3]{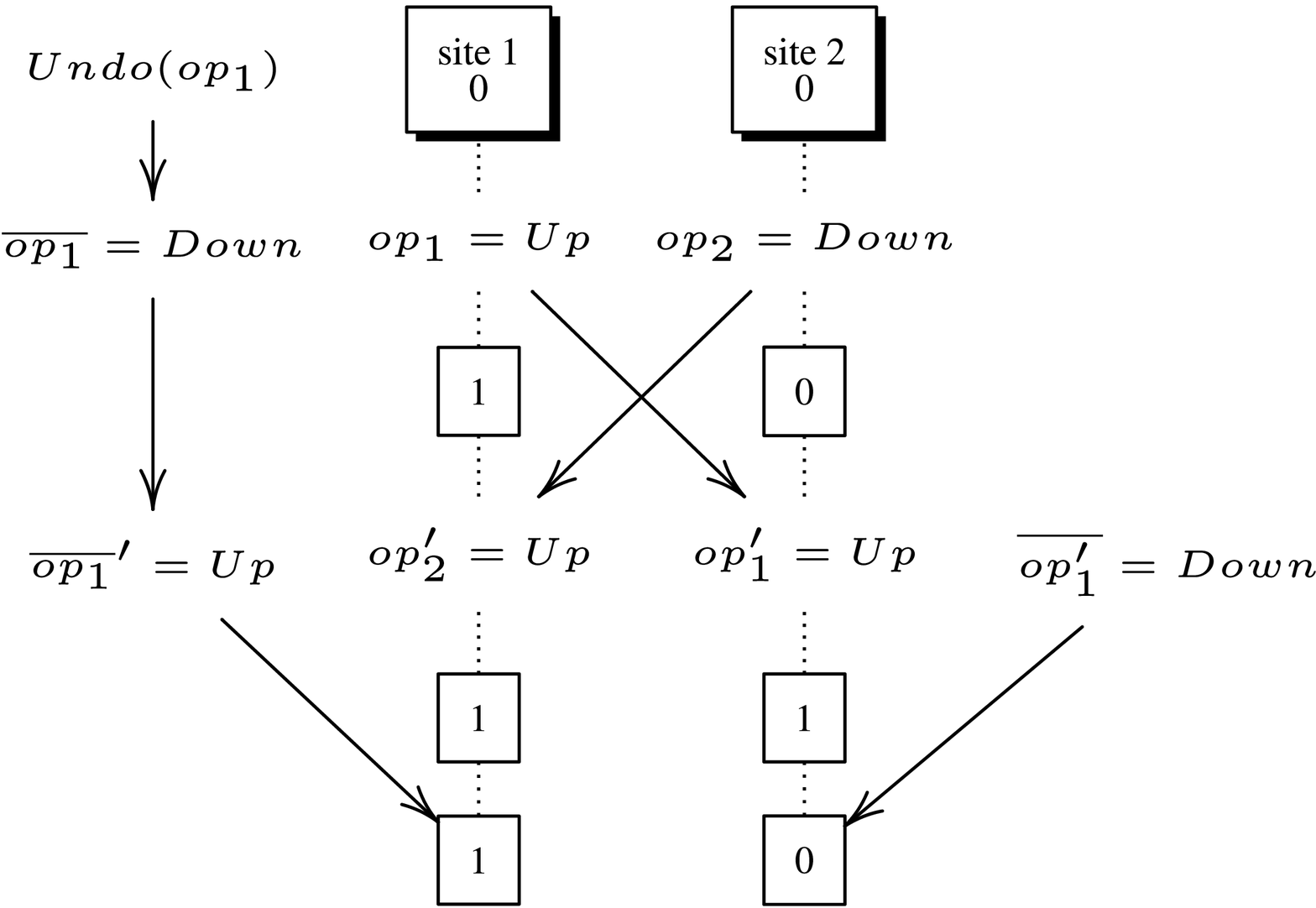}}
\begin{small}
\centerline{(b) $\mathbf{IP3}$ violation by the transformation function $IT_{bin}$.}
\end{small}
\end{minipage}
\caption{$\mathbf{IP3}$ property.}\label{fig:ip3}
\end{figure}

In general, $\mathbf{IP2}$ violation is discarded by placing the inverse of an undone operation  just after it in the log. The sequence $op \cdot \overline{op}$ is then marked in order to be ignored when transforming another operation against it. 
The violation of $\mathbf{IP3}$ cannot be avoided by such a mechanism and must be fulfilled by transformation functions in order to always ensure the data convergence.
To further illustrate inverse properties, we present the following examples:
 
\begin{example}\label{exmp:incDec}
Consider a shared integer register altered by two operations $Inc()$ and $Dec()$ which increments and decrements respectively the register state such as the one is the inverse of the other. A correct transformation function is defined as $IT(op_i,op_j)=op_i \mbox{ for all operations}\; op_i, op_j \mbox{ in } \{Inc(), Dec()\}$.  Note that operation $Inc()$ commutes with $Dec()$. Obviously, property $\mathbf{IP1}$ is satisfied since we have $  Inc() \cdot Dec() \equiv   Dec() \cdot Inc() \equiv \emptyset$.

Furthermore it is easy to verify that $\mathbf{IP2}$ and $\mathbf{IP3}$ are satisfied. On the one hand, $
IT(IT(op_1,op_2), \overline{op_2})= IT(op_1, \overline{op_2})=op_1.$
Thereby, showing that  $\mathbf{IP2}$ is indeed satisfied by the transformation function.
On the other hand, $
IT(\overline{op_1},IT(op_2,op_1))=IT(\overline{op_1},op_2)
=\overline{op_1}$. 
Since $\overline{IT(op_1,op_2)} = \overline{op_1}$,  
we  deduce that   indeed $IT(\overline{op_1},IT(op_2,op_1))= \overline{IT(op_1,op_2)}$ for all operations $op_1, op_2\in \{Inc(),Dec()\}$. Thus,  $\mathbf{IP3}$ is fulfilled. In Figure~\ref{fig:ip3}(a), we illustrate how $\mathbf{IP3}$ is preserved when undoing the operation $Inc()$ generated concurrently to $Dec()$. In this figure,  two sites  edit concurrently a shared integer. Initially, both sites have the state $0$. Site 1 increments the integer to get the state $1$ while site $2$ decrements it and gets the state $-1$. Every operation is integrated remotely to converge to the state $0$. Next, site 1 undoes the operation $op_1= Inc()$. For this,  $\overline{op_1}=Dec()$  is generated then transformed against the remote operation $op'_2=Dec()$ which leads to the final state $-1$. At site $2$, undoing $Inc()$ consists in generating its inverse $Dec()$ which leads to the same state as site $1$. Obviously, property $\mathbf{IP3}$ is preserved and both sites converges to the state $-1$.  Similarly, it is easy to check that $\mathbf{IP3}$ is also preserved if the operation $Dec()$ were undone.
\end{example}

\begin{example}\label{exmp:e14}
Consider again the shared binary register given in Example~\ref{exmp:upDown}.
Property $\mathbf{IP1}$ is violated since $0\cdot Down \cdot Up =1 \neq 0$. As for  property $\mathbf{IP2}$, it is violated since $IT(IT(Down, Down),Up)=Up\neq Down$. 
To illustrate the violation of $\mathbf{IP3}$, we consider  the Figure~\ref{fig:ip3}(b) where we show how to undo $op_1$ in the same situation depicted in Figure~\ref{fig:div}(b). Initially, both collaborating sites have a shared register set to   $0$. The sites perform concurrently then exchange the   operations $op_1=Up$ and $op_2=Down$ and converge to the final state $1$. Suppose now that the operation $op_1$ is undone at both sites.
At site $1$, undoing $op_1$ proceeds as follows:
\begin{inparaenum}[(i)]
\item  generate the  inverse of $op_1$ let $\overline{op_1}=Down$;
\item transform $\overline{op_1}$ against $op'_2$ which results in $IT(Up, Down)=Up$.
\end{inparaenum}  Thus, the final state after undoing $op_1$ is $1$ at site $1$. However, at site $2$, the execution of $\overline{op'_1}=Down$ at state $1$ produces the state $0$.
Consequently, both copies diverge due to the violation of $\mathbf{IP3}$  since $IT(\overline{op_1}, IT(op_2,op_1))\neq  \overline{IT(op_1,op_2)}$.
\end{example} 

Accordingly, it is easy to show by counterexamples that undoability is not always achieved even though the transformation function is correct. The question that arises here is how to define a transformation function that fulfills inverse properties? To answer this question, we propose in the following  to formalize the undoability problem as a CSP.

\section{Formalizing the Undoability Problem}\label{sec:formalProbStatement}
The undoablity problem  consists in investigating the existence of transformation functions satisfying both transformation and inverse
properties. In this section, we first provide a formal definition for collaborative objects and next, we formulate our undoability problem as a CSP.

\subsection{Consistent Collaborative Object (CCO)}
 
We suppose that there are $N$ sites (or users) collaborating on the same shared object replicated at each site.
Every site updates its local copy, executes the update immediately, then broadcasts it to other sites.
Before they are executed locally, remote updates are transformed against concurrent operations from the local log of the receiver site using the $IT$ function in order to integrate their effects. 

We formally define  a consistent collaborative object as follows:
\begin{definition}[Consistent Collaborative Object]\label{def:cco}
A \emph{Consistent Collaborative Object}  (CCO) is a triplet $C=\langle\mathcal{S}t,\mathcal{O}p,IT\rangle$ with:
\begin{itemize}[$\bullet$]
\item $\mathcal{S}t$ is a countable set of object \emph{states} (or the space state).\vspace{-.15cm}
\item $\mathcal{O}p$ is a countable set of \emph{primitive operations} executed by the user to \emph{modify} the object state, such that
each operation $op$ in $\mathcal{O}p$ has \emph{unique} and \emph{distinct  inverse} $\overline{op} \in \mathcal{O}p$
in such a way, applying $op$ followed by $\overline{op}$ has no effect.\vspace{-.15cm}
\item $IT : \mathcal{O}p\times \mathcal{O}p \rightarrow \mathcal{O}p$ is a \emph{correct} transformation function (\textit{i.e.},
$IT$ satisfies properties $\mathbf{TP1}$ and $\mathbf{TP2}$). 
\end{itemize}
 A CCO is of order $n$, denoted $n$-CCO, if the size of $\mathcal{O}p$ is equal to n.
\end{definition}

According to Definition~\ref{def:cco},  a CCO has no \emph{idle} operations (\textit{i.e.}, there is no $op\in \mathcal{O}p$ such that $st\cdot op = st$ for
any state $st$). Indeed,  when designing a shared object, a developer provides intuitively only operations that alter/modify the
object state. For her/him, it does not make sense to handle practically idle operations.  In addition to that, each operation has a unique and distinct (i.e. $op \neq \overline{op}$)
inverse, all operations have to satisfy the undo property $\mathbf{IP1}$ (see Definition~\ref{Def:IP1}).
Moreover,  the size of CCO is always even as each operation is different from its inverse. 
As seen in  previous examples, we can devise consistent objects (i.e. $\mathbf{TP1}$ and $\mathbf{TP2}$ are satisfied) 
without idle operations (see Examples~\ref{exmp:incDec} and~\ref{exmp:upDown}). 
We also exclude operations equal to their inverses (\emph{i.e}, $op=\overline{op}$) since they are not interesting in practice.

Note that Example~\ref{exmp:upDown} is not a CCO since $\mathbf{IP1}$ is violated while it is easy to prove that Example~\ref{exmp:incDec} is a CCO. 
 
A consistent collaborative object is said to be undoable if its transformation function verifies inverse properties $\mathbf{IP2}$ and $\mathbf{IP3}$ since $\mathbf{IP1}$ is assumed.
 
 \begin{definition}[Undoability]\label{def:undo}
A consistent collaborative object $C=\langle\mathcal{S}t,\mathcal{O}p,IT\rangle$ is \emph{undoable} iff its transformation
function $IT$ satisfies undo properties $\mathbf{IP2}$ and $\mathbf{IP3}$.
\end{definition}
 
In the sequel, all used objects are consistent collaborative objects (see Definition~\ref{def:cco}).

\subsection{CSP Statement}
Given a consistent collaborative object $C=\langle\mathcal{S}t,\mathcal{O}p,IT\rangle$,
our undoability problem consists in finding all transformation functions satisfying inverse
properties. However, this task  turns out to be a combinatorial problem. This is why, we propose to formalize the undoability problem as a CSP. 
Indeed, CSPs~\cite{CSP} are mathematical problems defined as a set of objects whose state must satisfy a number of constraints. 
They represent the entities in a problem as a homogeneous collection of finite constraints over variables, which is solved by constraint satisfaction methods.   CSPs  are solved in a reasonable time thanks to the combination of heuristics and combinatorial search methods.
Formally, a CSP is defined as follows:  

\begin{definition}[CSP]\label{def:csp}
A CSP is defined as a triple $\langle \mathcal{X},\mathcal{D},\mathcal{C} \rangle$, where:
\begin{itemize}
\item $\mathcal{X}=\{x_1,\ldots,x_n\}$ is a set of problem variables; \vspace{-1.5mm}
\item  $\mathcal{D}=\{\mathcal{D}_1,\ldots, \mathcal{D}_n\}$ is the set of   domain values for every the variable, \textit{i.e.} for every $k\in [1;n]$, $x_k \in \mathcal{D}_k$; and\vspace{-1.5mm} 
\item  $\mathcal{C}=\{\mathcal{C}_1,\ldots, \mathcal{C}_m\}$ is a set of constraints. 
 The  constraints may  be defined as    
\begin{inparaenum}[(i)]
\item arithmetic constraints such as $=$, $\neq$, $<$, $\leq$, $>$, $\geq$;
\item logical constraints  such as disjunction, implication, \textit{etc.}
\end{inparaenum}
\end{itemize} 
 An evaluation of the variables is a function from variables to values. A solution is an evaluation that satisfies all constraints from $\mathcal{C}$.

\end{definition}

Inspired by some famous CSP problems such as the eight-queen problem~\cite{NQueen,Bell20091}, we  formalize the  undoability problem  using CSP theory. Indeed, the undoability problem could be represented by  a square matrix where rows and columns refer to the operations  while their intersection  refers to the transformation result. 
In the following,  we discuss the ingredients of our  CSP model.

\medskip
\noindent\textbf{\textbf{The set of variables.}}
 $\mathcal{X}$ is the set of different possible  values taken by the operations to be transformed. Formally, given a CCO $C=\langle\mathcal{S}t,\mathcal{O}p,IT\rangle$ such that $\mathcal{O}p=\{op_1,\ldots,op_n\}$, then  $\mathcal{X}=\{IT(op_i,op_j)|$ $ op_i,\,op_j\in \mathcal{O}p \}$. %
For instance, if $\mathcal{O}p=\{op_1,op_2\}$, then $\mathcal{X}=\{IT(op_1,op_1), IT(op_1,op_2), IT(op_2,op_1), IT(op_2,op_2)\}$. Subsequently, a $n$-CCO has a set of variables  $\mathcal{X}$ of size  $n^2$.

\medskip
\noindent\textbf{\textbf{The domain.}} The domain of values is the set of values of each of the variables, \textit{i.e.} the transformation result of the $IT$ function for a couple of operations. Obviously,    we have $\mathcal{D}=\mathcal{O}p$. To simplify our model, we consider  $\Nat$ as the domain of operations. We represent the transformation function  $IT$ of a $n$-CCO by a square matrix of size $n^2$   such that operations corresponds to the indexes of rows and columns. The intersection of row $i$ with column $j$ is the evaluation of the transformation function $IT(i, j)$.  This representation has   $n^{n^{2}}$ different possible assignments in the search space which is too large.

\medskip
\noindent\textbf{\textbf{The set of constraints.}}
The constraints are the key component in 
expressing a problem as a CSP. They are the conditions to be satisfied by the $IT$ function so that an evaluation is true. Thus,  constraints are  $\mathbf{TP2}$,   $\mathbf{IP2}$ and $\mathbf{IP3}$. We exclude   properties  $\mathbf{TP1}$  and $\mathbf{IP1}$ since the former cannot be expressed by mathematical relations between the different variables from $\mathcal{X}$, while the latter is assumed.

Practically,    it is possible to find   useless solutions  while verifying both convergence and inverse properties. For instance, a correct transformation function may  just  undo the effect of the remote operation or that of the local operation against which it is transformed, as we will detail in  Example~\ref{ex:casBizarre}. Thus, after synchronizing all operations, all users will loose their updates which is far from being the objective of OT approach. 
To illustrate this, let us consider the following example:   
\begin{example}\label{ex:casBizarre}
Let $op_1 $ and $op_2$ be two  operations over square matrices of order 2 $M_{2,2}$, such that:\\
\begin{tabular}{l l}
\begin{tabular}{r r r l}
$op_1$ :  &$M_{2,2}$ &$\rightarrow$ &$M_{2,2}$\\
&$A$ &$\mapsto$ &$2 \times ^tA$\footnote{$^tA$ is the transpose of the matrix $A$.}\\
\end{tabular}&
\begin{tabular}{r r r l}
$op_2$ :&$M_{2,2}$ &$\rightarrow$ &$M_{2,2}$\\
&$A$ &$\mapsto$ &$ 2\times \begin{pmatrix}
a_{0,1}&a_{0,0} \\
a_{1,0}&a_{1,1}
\end{pmatrix}$\\ 
\end{tabular}
\end{tabular}\\
 
Consider the set of operations $\mathcal{O}p=\{op_1,op_2, op_3,op_4\}$ where $op_3$ and $op_4$ are the inverses of $op_2$ and $op_1$ respectively. The following correct transformation function may be defined over $\mathcal{O}p$:
\begin{inparaenum}[(i)]
\item $IT(op_1,op)=op_2$;  \item $IT(op_2,op)=op_1$;
\item $IT(op_3,op)=op_4$; and  \item $IT(op_4,op)=op_3$;
\end{inparaenum} where $op \in \mathcal{O}p$.
According to $\mathbf{TP1}$, the following equalities should be satisfied:\vspace{-.2cm} 
\begin{align}
             op_2 \cdot op_2&\equiv  op_1 \cdot op_1 \\
             \overline{op_2} \cdot op_1&\equiv  op_2 \cdot \overline{op_1} \\
             \overline{op_1} \cdot op_2&\equiv  op_1 \cdot \overline{op_2} \\
             \overline{op_1} \cdot \overline{op_1} &\equiv  \overline{op_2} \cdot \overline{op_2}  
        \end{align}
     
The above $IT$ function  satisfies each of these properties. Indeed, for every matrix $A$, we have $A\cdot op_1 \cdot op_1= 4  \times A$ and $A \cdot op_2 \cdot op_2 = 4 \times A$ which means equivalence (1) is satisfied. Similarly, equivalence (4) is satisfied.  
As for equivalence (2), it is correct since for every matrix $A \in M_{2,2}$, $\overline{op_2} \cdot op_1 = \frac{1}{2} \times op_2 \cdot op_1$ and $  op_2 \cdot \overline{op_1} =  op_2 \cdot \frac{1}{2} \times op_1$. Clearly, $\frac{1}{2} \times op_2 \cdot op_1 \equiv op_2 \cdot \frac{1}{2} \times op_1$.
Similarly,  equivalence (3) is correct. 

It is easy to prove the correctness of the previous transformation function. Nevertheless, such  a transformation does not make sense since it just undoes the effect of the performed operation when receiving a concurrent remote operation. For instance $IT(op_3, op_2)=op_4=\overline{op_2}$.  
\end{example}

 
Consequently, we propose to enhance our CSP model with 
the  constraints $\mathbf{C1}$ (see Definition~\ref{def:C1}) and $\mathbf{C2}$ (see Definition~\ref{def:C2}) in order to avoid undesirable $IT$ evaluations that  hide the advantage of OT approach (i.e. including the effect of concurrent operations).

Property $\mathbf{C1}$  forbids transforming an operation into its inverse:
\begin{definition}[Property $\mathbf{C1}$]\label{def:C1}
Given a CCO $C=\langle\mathcal{S}t,\mathcal{O}p,IT\rangle$ then for every operations $op_i$ and $op_j$ from $\mathcal{O}p$, it must be that $IT(op_i,op_j)\neq \overline{op_i}$.
\end{definition}

 As for property $\mathbf{C2}$, it discards $IT$ functions transforming an operation $op_1$ against $op_2$ to the inverse of $op_2$. 

\begin{definition}[Property $\mathbf{C2}$]\label{def:C2}
given a CCO $C=\langle\mathcal{S}t,\mathcal{O}p,IT\rangle$ then for every operations $op_i$ and $op_j$ from $\mathcal{O}p$, if $op_j\neq\overline{op_i}$ then $IT(op_i,op_j)\neq \overline{op_j}$.
\end{definition}

Accordingly, the final set of constraints is   $\mathcal{C}=\{\mathbf{TP2}, \mathbf{IP2}, \mathbf{IP3}, \mathbf{C1}, \mathbf{C2}\}$.


\section{Analysis of Transformation Patterns}

To obtain all the experimental results of the undoability problem \textit{i.e.} calculate all the evaluations of $IT$ with respect to our CSP model  
in a reasonable time, we have implemented a java prototype based on  the  Choco solver~\cite{choco}. Choco is a free and open source java library dedicated to constraint programming that allows describing combinatorial problems in the form of constraint satisfaction problems and solves them with constraint programming techniques. 

As we represent the transformation function by a square matrix, it is possible to have symmetric solutions (by rotation and reflection). To provide only effective solutions, we implemented a module that eliminates all symmetric solutions. 
%

In this section, we present how should be the transformation function so that undoability is correctly managed. 
In particular, we study whether commutativity is necessary and sufficient for undoablity or not. Our question stems from observing Examples $3$ and $4$: the shared integer register is undoable and its operations are commutative,
but the shared binary register is not   and its operations do not commute.

To answer the previous question, we begin by defining commutativity property and its implications on
transforming and undoing operations. Next, we analyze the output of our solver for CCOs of orders 2, 4 and 6 respectively.

\subsection{Commutativity Property}
 

We formally define the commutativity as follows.
\begin{definition}[Commutativity]\label{def:commutativity}
Two  operations  $op_1$ and $op_2$   commute iff  $  op_1\cdot op_2 \equiv  op_2 \cdot op_1$.
\end{definition}

In the following, we say that a set of operations $\mathcal{O}p$ is commutative if all of its operations are pairwise commutative.

Commutativity property given in Definition~\ref{def:commutativity} is strong in the sense that it enables us to reorder
any pair of operations whatever they are concurrent or causally dependent. Instead, in collaborative applications, we just need  to verify whether pairwise concurrent operations commute or not.
The impact of commutativity on $IT$ function  is shown in the following Theorem:

\begin{theorem}
\label{thr:concCommutativity}  
 For any pairwise concurrent operations $op_1, op_2\in \mathcal{O}p$, $op_1$ commute with $op_2$   iff   $IT(op_i,op_j)=op_i$, $i=1,2$.
\end{theorem}

\begin{proof*}
As the transformation function $IT$ is correct (see Definition~\ref{def:cco}), then $IT$ satisfies $\mathbf{TP1}$. That is, $  op_1\cdot IT(op_2,op_1) \equiv   op_2\cdot IT(op_1,op_2)$. Since, 
$  IT(op_2,op_1) =op_2$ and $ IT(op_1,op_2)= op_1$, we deduce from the previous equivalence that $op_2 \cdot op_1 \equiv op_1 \cdot op_2$. Consequently, $op_1$ commutes with $op_2$.
Moreover, if $\mathcal{O}p$ is commutative then for every two operations $op_1$ and $op_2$ from $\mathcal{O}p$, we have $op_1 \cdot op_2 \equiv op_2 \cdot op_1$. Consequently,   $IT(op_1,op_2)=op_1$ and $IT(op_2,op_1)=op_2$ according to $\mathbf{TP1}$.
Hence, for any pairwise concurrent operations $op_1, op_2\in \mathcal{O}p$, $op_1$ commutes with $op_2$   iff   $IT(op_i,op_j)=op_i$, $i=1,2$.
\end{proof*}

A natural follow-up question is how to define  a transformation function so that the collaborative
object is undoable and whether commutativity is necessary to achieve undoability or not?

First, we prove  that commutativity is sufficient for undoability. In other words,
we  show   that for any given consistent object $C=\langle\mathcal{S}t,\mathcal{O}p,IT\rangle$,   if   $ IT(op_i,op_j)=op_i$ for all concurrent operations $op_i$ and $op_j$ from $\mathcal{O}p$    then $C$ is undoable (see  Lemma~\ref{lemm:commRev}).
 

\begin{lemma}[Commutativity implies undoability]\label{lemm:commRev}
Given an object $C=\langle\mathcal{S}t,\mathcal{O}p,IT\rangle$, 
if $\mathcal{O}p$ is commutative then $C$ is undoable. 
\end{lemma}

\begin{proof*}
To prove that $C$ is undoable, we  have to verify that $\mathbf{IP2}$ and $\mathbf{IP3}$ properties are preserved.
Since $\mathcal{O}p$ is commutative then every operation is transformed to itself. Thus, for every two operations $op_i$ and $op_j$ from $\mathcal{O}p$,  we have
$
IT(IT(op_i,op_j),\overline{op_j})=IT(op_i,\overline{op_j})=op_i
$. 
 Then, $\mathbf{IP2}$  is satisfied.
As for $\mathbf{IP3}$, it is satisfied since,
$
IT(\overline{op_i}, IT(op_j,op_i)) =  IT(\overline{op_i}, op_j) = \overline{op_i}= \overline{IT(op_i,op_j)}
$.
\end{proof*}

In the following, we discuss the solutions provided by our prototype for orders $2$, $4$ and $6$ and see whether they commute or not. 
\subsection{CCO of order 2}
 To discuss the correct evaluations of the transformation function $IT$ in the case of a $2$-CCO, we consider  $\mathcal{O}p=\{op_1, op_2\}$ such that $\overline{op_1}=op_2$. 
When enforcing the set of constraints, only one solution was provided by our solver (see Figure~\ref{fig:2CCO}).
\begin{table*}[htbp]

      \centering \begin{small} 
      \begin{tabular}{|c|c|c|}
\hline
$IT$&  $op_1$ &   $op_2$  \\ \hline
$op_1$& $op_1$ & $op_1$\\ \hline 
$op_2$ &  $op_2$& $op_2$\\ \hline 
\end{tabular}
\end{small}
   \captionof{figure}{Output of the 2-CCO problem.}\label{fig:2CCO}
\end{table*}
\vspace{-.4cm} 
 This output shows that an undoable CCO of order $2$  requires a transformation function verifying  $IT(op_i, op_j)=op_i$  for every pairwise operations $op_i$ and $op_j$ from 
$\mathcal{O}p$ ($i.e$ every operation is transformed to itself) thereby $\mathcal{O}p$ commute as stated in Theorem~\ref{thr:concCommutativity}. Accordingly,   the commutativity is necessary to correctly undo concurrent operations in the case of $2$-CCOs.

\subsection{CCO of order 4}
Figure~\ref{fig:newCSP4} shows the output of our solver in the case of $4$-CCOs. For this experiment, we have considered a set of four operations $\mathcal{O}p=\{op_1, op_2, op_3, op_4\}$ such that $\overline{op_1}=op_4$ and  $\overline{op_2}=op_3$. Similarly to $2$-CCOs, commutativity is  necessary to achieve undoability  since every operation is transformed to itself (see Theorem~\ref{thr:concCommutativity}).
\begin{table*}[htbp]  
		\centering \begin{small}  
      \begin{tabular}{|c|c|c|c|c|}
\hline
$IT$&  $op_1$ &   $op_2$& $op_3$ &   $op_4$  \\ \hline
$op_1$& $op_1$& $op_1$ &$op_1$& $op_1$\\ \hline 
$op_2$& $op_2$& $op_2$ &$op_2$& $op_2$\\ \hline 
$op_3$& $op_3$& $op_3$ &$op_3$& $op_3$\\ \hline
$op_4$& $op_4$& $op_4$ &$op_4$& $op_4$\\ \hline
\end{tabular}
     \end{small} 
   \captionof{figure}{Output of the  $4$-CCO problem.}\label{fig:newCSP4}
\end{table*}
\vspace{-.4cm}

\subsection{CCO of order 6}
We discuss here the transformation functions provided by our solver 
for CCOs of order 6. We have considered that  $\mathcal{O}p=\{op_1, op_2, op_3, op_4, op_5, op_6\}$ such that $\overline{op_1}=op_6$, $\overline{op_2}=op_5$ and  $\overline{op_3}=op_4$. Figure~\ref{fig:6CCO} shows that three solutions are possible to attain undoability.
\begin{table*}[htbp]
\begin{scriptsize}
   \hfill\hfill
   \begin{minipage}[t]{0.33\linewidth}
      \centerline{(1) Solution 1}
      \centering 
   \begin{tabular}{|c|c|c|c|c|c|c|}
\hline
$IT$&  $op_1$ &   $op_2$& $op_3$ &   $op_4$  & $op_5$ & $op_6$ \\ \hline
$op_1$& $op_1$& $op_4$ &$op_1$& $op_1$ &$op_4$ &$op_1$ \\ \hline 
$op_2$& $op_2$& $op_2$ &$op_2$& $op_2$ & $op_2$  & $op_2$ \\ \hline 
$op_3$& $op_3$& $op_6$ &$op_3$& $op_3$ &$op_6$ & $op_3$\\ \hline
$op_4$& $op_4$& $op_1$ &$op_4$& $op_4$ &$op_1$ & $op_4$ \\ \hline
$op_5$& $op_5$& $op_5$ &$op_5$& $op_5$ & $op_5$& $op_5$ \\ \hline
$op_6$& $op_6$& $op_3$ &$op_6$& $op_6$ & $op_3$ & $op_6$\\ \hline
\end{tabular}
   \end{minipage}
   \hfill\hfill
   \begin{minipage}[t]{0.5\linewidth}
      \centerline{(2)  Solution 2}
      \centering      
      \begin{tabular}{|c|c|c|c|c|c|c|}
\hline
$IT$&  $op_1$ &   $op_2$& $op_3$ &   $op_4$& $op_5$ & $op_6$   \\ \hline
$op_1$& $op_1$& $op_4$ &$op_1$& $op_1$ & $op_3$& $op_1$\\ \hline 
$op_2$& $op_2$& $op_2$ &$op_2$& $op_2$ & $op_2$ & $op_2$ \\ \hline 
$op_3$& $op_3$& $op_1$ &$op_3$& $op_3$ &$op_6$ & $op_3$ \\ \hline
$op_4$& $op_4$& $op_6$ &$op_4$& $op_4$ & $op_1$& $op_4$\\ \hline
$op_5$& $op_5$& $op_5$ &$op_5$& $op_5$ & $op_5$& $op_5$ \\ \hline
$op_6$& $op_6$& $op_3$ &$op_6$& $op_6$ & $op_4$& $op_6$ \\ \hline
\end{tabular}
   \end{minipage}
   \hfill \hfill
	 \begin{minipage}[t]{0.99\linewidth}
      \centerline{(3) Solution 3}
      \centering 
   \begin{tabular}{|c|c|c|c|c|c|c|}
\hline
$IT$&  $op_1$ &   $op_2$& $op_3$ &   $op_4$& $op_5$ & $op_6$  \\ \hline
$op_1$& $op_1$& $op_1$ &$op_1$& $op_1$ & $op_1$& $op_1$ \\ \hline 
$op_2$& $op_2$& $op_2$ &$op_2$& $op_2$ & $op_2$& $op_2$ \\ \hline 
$op_3$& $op_3$& $op_3$ &$op_3$& $op_3$ & $op_3$ & $op_3$ \\ \hline
$op_4$& $op_4$& $op_4$ &$op_4$& $op_4$ & $op_4$ & $op_4$ \\ \hline
$op_5$& $op_5$& $op_5$ &$op_5$& $op_5$ & $op_5$& $op_5$ \\ \hline
$op_6$& $op_6$& $op_6$ &$op_6$& $op_6$ & $op_6$& $op_6$ \\ \hline
\end{tabular}
   \end{minipage}
	\end{scriptsize}
   \captionof{figure}{Output of the 6-CCO problem.}\label{fig:6CCO}
\end{table*}
Thus an undoable 6-CCO is not necessarily commutative. Indeed, among the three  solutions provided by our solver only the last one commutes according to Theorem~\ref{thr:concCommutativity}. However, a very important observation that can be made is: $4$ operations from the operations set are transformed at least $4$ times to themselves and $2$ others are always transformed to themselves. The analysis of both solutions 1 and 2 shows each solution is formed by a commutative CCO of order $4$ and another commutative CCO of order $2$ such that the transformation inter-CCOs (transformation between $4$-CCO and $2$-CCO) does not commute. 

\section{Discussion}   

Our previous study proves that commutativity is closely related to undoability while the OT approach was proposed to go beyond commutativity.  
Indeed, CCOs of order $2$ and $4$ are undoable if and only if  they commute as stated in the following theorem: 
 \begin{theorem}\label{thr:Th2-4} 
Commutativity is necessary and sufficient to achieve undoablity for CCOs of order $n\leq 4$.
\end{theorem} 
\begin{proof*}
The experimental results obtained by executing the solver for CCOs of order $2$ and $4$ show   that commutativity is necessary to achieve undoability. Since commutativity is also sufficient to achieve undoability (see Lemma~\ref{lemm:commRev}), we deduce that commutativity is equivalent to undoablilty for CCOs of orders $2$ and $4$.
\end{proof*}

However,  commutativity is sufficient but not necessary to achieve undoablitiy in the case of CCOs of order $n\geq 6$.
Indeed, our solver provides three correct transformation functions that are undoable where only  one is commutative according to Theorem~\ref{thr:concCommutativity}. The two others do not commute but consist of two sub-sets of commutative operations. Accordingly,  a $6$-CCO is formed by two  intra-commutative sub-CCOs  such that   the transformation inter both CCOs does not commute. 
The analysis of the output presented in Figure~\ref{fig:6CCO} shows that the set of operations $\mathcal{O}p$ of any undoable 6-CCO is:
\begin{enumerate}
\item either commutative, \textit{i.e.} $IT(op_i, op_j)= op_i$ for every  pair of operations $op_i$, $op_j$ in $\mathcal{O}p$; 
\item   or the union of two sub-sets $\mathcal{O}p_1$ and $\mathcal{O}p_2$ such that:
$\mathcal{O}p_1$ and $\mathcal{O}p_2$ are commutative of size 2 and 4 respectively (\textit{i.e.}  
$IT(op_i, op_j)=op_i$, for every pair of operations $(op_i, op_j) \in \mathcal{O}p_{x}\times \mathcal{O}p_{x}$, $x\in\{1,2\}$). The transformation inter CCOs is summarized as follows: 
\begin{enumerate}
\item for every pair of operations $(op_i, op_j) \in \mathcal{O}p_{1}\times \mathcal{O}p_{2}$, $IT(op_i,op_j)=op_i$
\item for every pair of operations $(op_i, op_j) \in \mathcal{O}p_{2}\times \mathcal{O}p_{1}$,
\begin{enumerate}
\item  either $IT(op_i, op_j)=IT(op_i, \overline{op_j})$;
\item or $IT(op_i, op_j)=\overline{IT(op_i, \overline{op_j})}$
\end{enumerate}
\end{enumerate}
\end{enumerate}

The solutions produced by our solver in the case of $8$-CCOs validate the previous observation and follow the patterns found above. However, due to space limit, we cannot present and discuss these solutions.  We strongly believe that the patterns found for $6$-CCOs  may be generalized  by induction  on the CCO's order. 


Moreover, our experiments provide a small number of solutions which greatly simplifies the study of the undoability problem. Indeed, our set of constraints  considerably reduces the number of correct evaluations for transforming concurrent operations which  saves time and effort when designing a concurrent application. For instance, a $6$-CCO normally generates $6^{6^2}$ transformation functions while we  only obtain $3$ patterns. 
This would be very useful for collaborative applications designers.

To summarize, this work proves that there is only one possible way of transforming concurrent operations for CCOs of order $2$ and $4$ to ensure they are undoable. This unique solution consists in transforming each operation to itself thus the commutativity is necessary and sufficient to achieve undoablility. Otherwise,   commutativity is only sufficient. Furthermore, an undoable CCO of order $n\geq 6$ is the union of two intra-commutative sub-CCOs which allows devising generic transformation patterns useful for the design of collaborative applications.  Yet OT approach was proposed to go beyond commutativity, this work shows that commutativity somehow impacts on undoability.

\section{Related Work}
Several works proposed undo capability for   collaborative editors.
The majority of these solutions are based on log usage in order to store operations and recover earlier states.

\emph{\textbf{Swap then undo}}~\cite{Atul94} was the first selective undo. It consists on  placing the selected
operation in the end of the history by swapping then executing its inverse.
Unfortunately, this solution does not allow to undo any operation since it is not always possible to swap operations in the log.   To avoid this issue, authors defined the boolean function $conflict()$  that aborts the undo procedure in conflicting situations.

\emph{\textbf{Undo/Redo}}~\cite{Ressel99} was proposed to overcome the conflict problem. It consists in undoing all the operations in the inverse chronological
order.
However, it is expensive since it requires to perform many steps and does
not allow undo in all cases since an operation may  not be undoable.

The approach of \textbf{Ferri\'e}~\cite{Ferrie04}   has a quadratic complexity and is based on  the transformation functions of the algorithm SOCT2~\cite{suleiman97} that violates convergence properties~\cite{ImineMOR03, Imi06}.

\textbf{\emph{UNO}}~\cite{WEISS:2008} consists in generating a new operation having the inverse effect of the operation to be undone. Although it has a linear complexity,   this solution only fits applications based on TTF~\cite{Oster06}   where characters are not effectively deleted from the document. Moreover,   the correctness proof of UNO assumes the intention preservation which is not proved formally~\cite{BinShao10}.

Both \textbf{\emph{ANYUNDO-X}}~\cite{Sun02}  and \textbf{\emph{COT}}~\cite{SunS09} support integrated Do and selective Undo and allow the undo
of any operation while solving the known undo problematic. However, they both have an exponential complexity and are based on avoiding some  inverse properties (namely $\mathbf{IP2}$ and $\mathbf{IP3}$) instead of fulfilling them.
In COT, a contextual relation is introduced to illustrate the relation between an operation, its inverse and the transformed intermediates forms of the inverse. The time complexity is also exponential in the log size.
The difference between ANYUNDO~\cite{Sun02} and COT~\cite{SunS09} is that the latter discuss the undo in the case of causally dependent operations and not only concurrent ones. 

Finally, the \textbf{\emph{ABTU}} algorithm~\cite{BinShao10}   proposes an undo solution     basing on the transformation algorithm ABT~\cite{LiLi10}.
Even though the proposed algorithm has a linear complexity, it  does not allow to undo any operation since undo is aborted in some cases.
The transformation algorithm ABT is based on   \textit{effect relation} allowing to order document updates in the log. Consequently, all updates are ordered according to their effect relation on the shared document state.
 Authors assume that this relation ensures convergence. However, this algorithm diverge in some cases.

\section{Conclusion and Future Work}
In this paper, we have presented a formal model for the undoability problem. Indeed, we have shown how to formulate the undoability problem as a CSP. Thus, it is possible to compute all correct transformation functions that achieve convergence and undoability  using a CSP solver. Our experiments showed that  undoability for CCOs of order $2$ and $4$ is  achieved if and only if the operations commute  which considerably simplifies the design of collaborative objects. However, for all CCOs of order $n\geq 6$, it is possible to define multiple transformation functions to achieve undoability. Fortunately, we have shown that these solutions are either commutative or formed by sub commutative CCOs. In future work,  we will deeply investigate in the transformation functions provided by our undoability solver in order to generalize the transformation patterns defined for 6-CCOs which allows to devise a generic transformation framework for finite and arbitrary set of operations. Such framework will be very useful for collaborative applications designers since it guarantees the correctness and undoability for any given solution. Furthermore, we will relax property $\mathbf{IP2}$ by providing alternative constraints since it is   always discarded by designers instead of being fulfilled.

\bibliographystyle{eptcs}
\bibliography{mybib}
\end{document}